\documentclass[preprint2]{aastex}

\newcommand\KAB{K_{\rm AB}}
\newcommand\Msun{M_\odot}
\newcommand\Ha{{\rm H}\alpha}
\newcommand\Hb{{\rm H}\beta}
\begin{document}

\title{
Star Formation Rates and Metallicities of $K$-selected Star Forming
Galaxies at $z \sim 2$\altaffilmark{1}
}

\author{
Masao Hayashi \altaffilmark{2},
Kentaro Motohara \altaffilmark{3},
Kazuhiro Shimasaku \altaffilmark{4},
Masato Onodera \altaffilmark{5,8},
Yuka Katsuno Uchimoto \altaffilmark{3},
Nobunari Kashikawa \altaffilmark{6},
Makiko Yoshida \altaffilmark{2},
Sadanori Okamura \altaffilmark{4},
Chun Ly \altaffilmark{7},
and
Matthew A. Malkan \altaffilmark{7}
}

\email{hayashi@astron.s.u-tokyo.ac.jp}

\altaffiltext{1}{
Based on data collected at Subaru Telescope, which is operated by
the National Astronomical Observatory of Japan.
Use of the UKIRT 3.8-m telescope for the observations is supported by NAOJ.}
\altaffiltext{2}{
Department of Astronomy, Graduate School of Science,
University of Tokyo, Tokyo 113-0033, Japan}
\altaffiltext{3}{
Institute of Astronomy, Graduate School of Science, University of
Tokyo, Mitaka, Tokyo 181-0015, Japan}
\altaffiltext{4}{
Department of Astronomy and Research Center for the Early Universe,
Graduate School of Science,
University of Tokyo, Tokyo 113-0033, Japan}
\altaffiltext{5}{
Department of Astronomy, Yonsei University, Sinchon-dong 134,
Seodaemun-gu, Seoul, Korea}
\altaffiltext{6}{
Optical and Infrared Astronomy Division, National Astronomical
Observatory, Mitaka, Tokyo 181-8588, Japan}
\altaffiltext{7}{
Department of Physics and Astronomy, University of California at Los
Angeles, P. O. Box 951547,  Los Angeles, CA 90095-1547, USA}
\altaffiltext{8}{
Service d'Astrophysique, CEA Saclay, Orme des Merisiers, 91191
Gif-sur-Yvette Cedex, France}

\begin{abstract}
 We present spectroscopy of 15 star-forming BzK galaxies
 (sBzKs) with $\KAB \lesssim 23$ in the Subaru Deep Field, for which
 $\Ha$ and some other emission lines are detected in 0.9 to 2.3
 $\micron$ spectra with a resolution of $R$=500. 
 Using $\Ha$ luminosities, we obtain star formation rates (SFRs), and
 then specific SFRs (SSFRs) dividing SFRs by stellar masses, 
 which are derived from SED fitting to $BVRi'z'K$ photometry. 
 It is found that sBzKs with higher stellar masses have larger SFRs. 
 A negative correlation is seen between stellar mass and SSFR,  
 which is consistent with the previous results for $z \sim 2$
 galaxies. This implies that a larger growth of stellar mass occurs in
 less massive galaxies.  
 In addition, gas-phase oxygen abundances, 12+log(O/H), are
 derived from the ratio of [\ion{N}{2}]($\lambda6584$) to $\Ha$ using
 the N2 index method. We have found a correlation between stellar mass
 and oxygen abundance in the sense that more massive sBzKs tend to be
 more metal rich, which is qualitatively consistent with the relation
 for UV-selected $z \sim 2$ galaxies. However, the metallicity of the
 sBzKs is $\sim 0.2$ dex higher than that of UV-selected galaxies with
 similar stellar masses, which is significant considering the small 
 uncertainties.   
 The sBzKs in our sample have redder $R-K$ colors than the UV-selected
 galaxies.  
 This galaxy color-dependence in the oxygen abundance may be caused by
 older or dustier galaxies having higher metallicities at $z \sim 2$.
\end{abstract}


\section{Introduction}
Recent studies suggest that the era of $z \sim 2$ is a turning point
in galaxy formation and evolution. The cosmic star formation
rate (SFR) begins to drop at $z \sim 1-2$ from a flat plateau at higher
redshifts \citep{dic03,fon03,ly07}. Significant evolution of the Hubble
sequence occurred at $z \sim 1-2$ \citep{kaj01}.   
It is also found that the number density of QSOs has a peak at 
$z \sim 2$ \citep{ric06}. These facts suggest that dramatic changes in
the galaxy population occurred at $z \sim 2$, which are important to 
study in further detail. 

Galaxy properties are determined from spectral
characteristics such as emission lines, absorption lines and strong
continuum breaks.  
At $z \sim 2$, the strongest of these features fall in the
near-ultraviolet (NUV) and near-infrared (NIR) wavebands.  
NUV and NIR observations are thus essential to reveal detailed
properties of $z \sim 2$ galaxies, but these observations encounter many 
difficulties, including poor sky transparency except for some
atmospheric windows, and strong OH airglow emission lines and thermal emission
from the atmosphere.
This is why the era of $z \sim 2$ had been called a
{\lq}redshift desert{\rq} until recently.  
However, recent advances in wide-field imaging in the NUV and NIR, along
with multi-object spectrographs, are beginning to fill in this
{\lq}desert{\rq} (e.g., Subaru Deep Field (SDF) survey, GOODS-North and
South, UKIDSS, GMASS, and MUSYC 
\citep{mccarthy99,red06b,erb06c,hay07,dad07,gra07,lan07,hal08,kri08,ly08}).  

A photometric selection technique using $B-z$ and $z-K$ colors has been
proposed to choose an unbiased sample of $z \sim 2$ galaxies (BzK
galaxies: \citet{dad04}). Using these colors to find galaxies with
a Balmer break between $z$ and $K$, the selected
galaxies are then classified into star forming galaxies (sBzKs) and
passively evolving galaxies (pBzKs). 
Thanks to wide and deep imaging surveys in the NIR as well as the optical, 
large new BzK samples are enabling us to measure their statistical
properties, such as clustering strength \citep{kon06,hay07,bla08}. 
The multiwavelength data also show that sBzKs are very actively
forming stars\citep{dan06,dad07}. On the other hand, 
due to lack of spectroscopic observations, there are relatively few
accurately known redshifts, which are needed   
to derive accurate stellar masses and star formation rates
from SED fitting to the multiwavelength data. Also, their physical
properties such as metallicity are still not well-known.

More spectroscopy has recently become available for 
$z \sim 2$ galaxies selected by their UV colors.
\citet{erb06a,erb06b} presented a mass-metallicity ($M$-$Z$) relation
and SFRs for UV-selected galaxies at $z \sim 2$ using 114 NIR spectra. 
They found a $M$-$Z$ correlation at $z \sim 2$ as
there is in the local universe, with the more massive galaxies being
more metal rich. Although the correlation has a similar slope to the
local one, it is shifted to lower metallicity by $\sim 0.3$ dex.
They have also found that SFRs derived from $\Ha$ luminosities are in
good agreement with those from de-reddened rest-UV luminosities, with an
average dust-corrected SFR of $\sim 30 \Msun {\rm yr^{-1}}$. 

However, the rest-UV color selection misses about half of $z \sim 2$
galaxies, which have a large amount of dust extinction \citep{kon06}. 
The spectroscopic properties of this dusty population have yet not been
clearly revealed. 
$K$-limited samples are better suited than UV-limited ones to
investigate the relations between stellar mass and other fundamental
properties at $z \sim 2$, because they approximately correspond to
samples limited by stellar mass, without excluding galaxies with large
dust extinction.  
\citet{kri08} carried out a NIR spectroscopic survey for $K$-bright
galaxies at $z \sim 2.3$. They suggest that studies with broadband
photometric data alone may overestimate the number of massive galaxies
at $z \sim 2$, and underestimate the evolution of the stellar mass
density, again indicating the importance of spectroscopy.

In this paper, we present NIR spectroscopy for 44 BzK galaxies in the
SDF, and their spectroscopic properties.  
The structure of this paper is as follows.
The photometric and spectroscopic data are described in \S
\ref{sec;data}. In \S \ref{sec;analysis}, we examine the properties of
the detected emission lines, including spectroscopic redshifts and
fluxes. We then investigate active galactic nuclei (AGN) 
contamination in our spectroscopic sample in \S \ref{sec;agn}.
In \S \ref{sec;discussions}, physical properties are derived from the
emission lines, and then discussed.   
A summary is given in \S \ref{sec;conclusion}. 
Throughout this paper, magnitudes are in the AB system, 
and we adopt cosmological parameters of $h=0.7$, 
$\Omega_{m0}=0.3$ and $\Omega_{\Lambda 0}=0.7$.

\section{Data}
\label{sec;data}

\subsection{BzK galaxies in the SDF}
\label{sec;BzKs}
Here we briefly describe the imaging data and sample selection of BzKs
in the SDF, which are almost the same as those in \citet{hay07}. 
The SDF is a large ``blank'' field centered on ($13^{\rm h} 24^{\rm m} 38\fs9$,
$+27\arcdeg29\arcmin25\farcs9$; J2000), a rectangle of 
$29\farcm7 \times 36\farcm7$, with
deep multiwavelength data from the NUV to
mid-IR (MIR) wavebands. Among these data, the optical and the NIR data
are mainly used in this study. The optical data, $B$, $V$, $R$, $i'$,
$z'$, NB816, and NB921, were obtained with the Suprime-Cam on the Subaru
telescope for the Subaru Deep Field Project \citep{kas04}, and then the
NIR data, $J$ and $K$, were obtained with the wide field camera (WFCAM)
on the United Kingdom Infra-Red Telescope (UKIRT) (Motohara et al. in
preparation).   
The NIR data available to this study were only what was obtained on 2005
April 14 and 15, which limited the overlapping area with
the optical data to about $40\%$ of the whole SDF \citep{hay07}. 
The $3\sigma$ limiting magnitudes in a $2\arcsec$-diameter aperture
are 28.45 magnitude for $B$, 26.62 for $z'$, and 24.05 for $K$. The
optical images have been convolved to a seeing size of $1\farcs14$,
which is the value for $K$. 

MIR data, 3.6 -- 8.0 $\micron$, were obtained with the InfraRed Array
Camera (IRAC) on the Spitzer space telescope. Since the point spread
functions in the MIR data are much broader than those in optical and
NIR data, some galaxy images are blended with nearby objects. 
For these, we had to remove the contaminating flux contributions from 
close objects in order to do accurate photometry. 
To do this we use GALFIT \citep{pen02} to
deblend images with adjacent objects (Hayashi et al. in
preparation), and accurately measure the MIR fluxes of objects. 
The $3\sigma$ limiting magnitudes are 23.60 for $3.6 \micron$, 23.32 for
$4.5 \micron$, 21.65 for $5.8 \micron$, and 20.99 for $8.0 \micron$,
which are measured with $4.4\arcsec$, $4.4\arcsec$, $5.0\arcsec$, and
$6.4\arcsec$ diameter apertures, respectively.    

We applied the $BzK$ color selection proposed by \citet{dad04} to the
$B$, $z'$, and $K$ data in the 180 arcmin$^2$ of the SDF which have good 
detection uniformity in all three passbands. In selecting BzK galaxies,
deep optical data are crucial, because objects without $B$ or $z'$
detections can take arbitrary positions on the $BzK$ color diagram, with
almost all of them left unclassified. We selected the BzKs with
great care for the optically undetected objects (See also
\citet{hay07}). We obtained samples of 1092 sBzKs and 56 pBzKs with 
$K < 23.2$, corresponding to the $5\sigma$ limiting total magnitude.

\subsection{Spectroscopic observation}
\label{sec;observation}
NIR spectroscopy of the BzKs in the SDF was carried out with the
Multi-Object InfraRed Camera and Spectrograph \citep[MOIRCS;][]{ich06}
on the Subaru telescope on 2007 May 3 and 4. MOIRCS has a wide field
of view of $4\arcmin \times 7\arcmin$, and its $zJ500$ and $HK500$
grisms provide spectra with a resolution of $R$=500, covering wavelengths
of 0.9 -- 1.8 $\micron$ and 1.3 -- 2.5 $\micron$, respectively.
The dispersions of the grisms are 5.57 and 7.72 \AA/pixel,
respectively. 

\begin{figure}[tb]
 \begin{center}
  \plotone{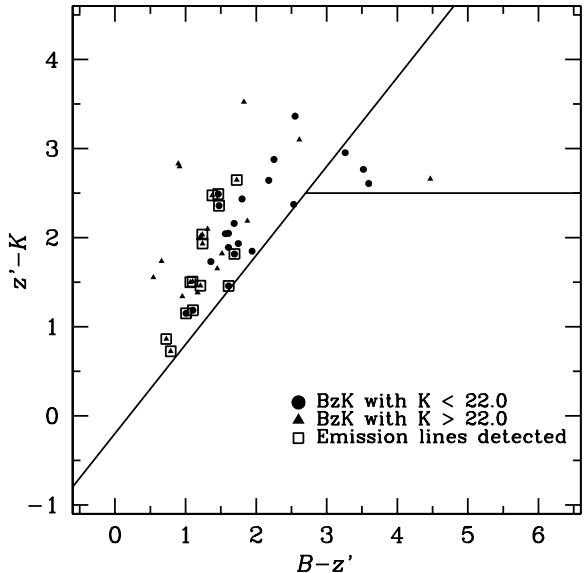}
  \caption{The colors of 44 targets. The filled circles are BzKs with
  $K<22.0$, and the filled triangles are those with $K>22.0$. BzKs with
  emission lines detected are squares.  
           }
  \label{fig;color_target}
 \end{center}
\end{figure}

Forty-four spectroscopic targets were selected from the samples of 1092 sBzKs
and 56 pBzKs. We gave priority to $K<22.0$ BzKs in our samples for 
slitmask design, and then $K>22.0$ BzKs were allocated to the blank
remaining slits. We used two slitmasks to cover the 44 BzKs, that is, each
slitmask covers 22 BzKs.   
Each slit has a width of $0\farcs7$ with a length of 10 -- 12
$\arcsec$ according to the space between the slits.
Among the 44 galaxies, 40 are classified as sBzKs and 4 as pBzKs. 
Half of them are fainter than $K=22.0$. 
Figure \ref{fig;color_target} shows the colors of the 44 BzKs. 

We observed them consecutively at two positions (A and B) on the slit,
using the $zJ500$ and $HK500$ grisms for one slitmask per night 
with single exposure times of 600 or 900 sec. 
The total on-source integration time was then 2 -- 2.6 hours for both
grisms.   
During the two nights, the weather was good, and the seeing was as
good as $\sim 0\farcs4$ in $K$, except for the end of May 4, when it was
$\gtrsim 1\farcs0$.  
We also observed A0 stars with $V \sim 9.0$ magnitude at different air
masses with both grisms on each night to correct for the telluric 
absorption and the instrumental efficiency. Before and after the
observations, dome flats with lamps on and off, and Th-Ar lamp
data were obtained for wavelength calibration. 

\setcounter{footnote}{8}

\subsection{Data Reduction}
\label{sec;reduction}
Data reduction is done with standard IRAF%
\footnote{IRAF is
distributed by the National Optical Astronomy Observatories, which are
operated by the Association of Universities for 
Research in Astronomy, Inc., under cooperative agreement with the
National Science Foundation.}
procedures.
First, we create A -- B frames from successive frames observed at
the two positions. The A -- B frames are then divided by a flatfield image, 
and bad pixels and cosmic rays are removed. 
The flat image is created by taking difference between the dome
flat frames with the lamps on and off, and normalizing it. 
Next, distortion is corrected using calibration data files provided by
the MOIRCS instrument team.    
Wavelength calibration is done with only the OH airglow lines for the
$HK$ data, since there are numbers of the strong OH lines in the
wavelengths longer than $1.1 \micron$. On the other hand, for the $zJ$
data, Th-Ar data are also used for wavelength calibration at
wavelengths shorter than $1.1 \micron$.  
The wavelength uncertainties in the strong lines are less than 15\AA.
  
Residual sky subtraction is then carried out, since only the A -- B
procedure may not completely remove the sky background due to its
time variation.   
The telluric absorption and the instrumental efficiency are corrected
using the spectra of A0 stars and the stellar spectral library given by
\citet{pic98}. 
Finally, all the frames for each galaxy are added and 1D spectra are
extracted combining 5 pixels along a slit.

Flux calibration of the spectra is done using the
photometry from the imaging data in $z'$ and $K$.   
The $1\sigma$ error of the spectrum is estimated from the standard
deviation in the sky region of the 2D spectrum.

\section{Measurement of emission lines}
\label{sec;analysis}
\subsection{Detection of emission lines}
\begin{figure*}[t]
 \begin{center}
  \epsscale{1.4}
  \plotone{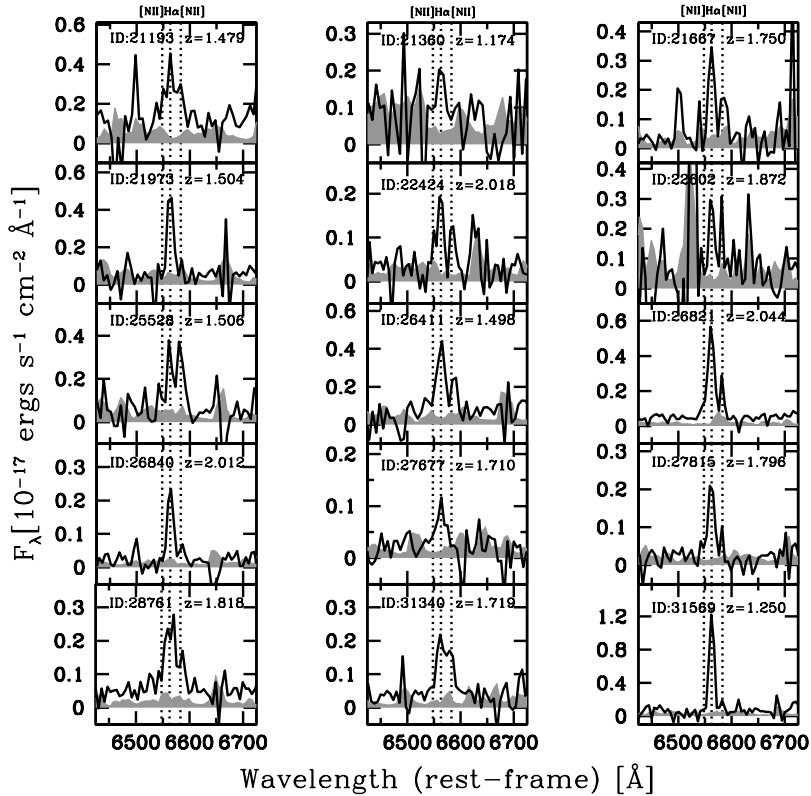}
  \caption{The spectra of $\Ha$ and [\ion{N}{2}] emission lines we have
  detected. These spectra have a binned dispersion of $\sim 16$ \AA.
  The solid lines show the spectra and the shaded regions in gray are
  the $1\sigma$ sky noise. We regard lines with a peak flux larger than 
  the $3\sigma$ sky noise as a detection using unbinned 1D spectra.
  Labeled redshifts are derived from $\Ha$ lines.
           }
  \epsscale{1.0}
  \label{fig;spectra-a}
 \end{center}
\end{figure*}

\begin{figure}[bth]
 \begin{center}
  \epsscale{1.1}
  \plotone{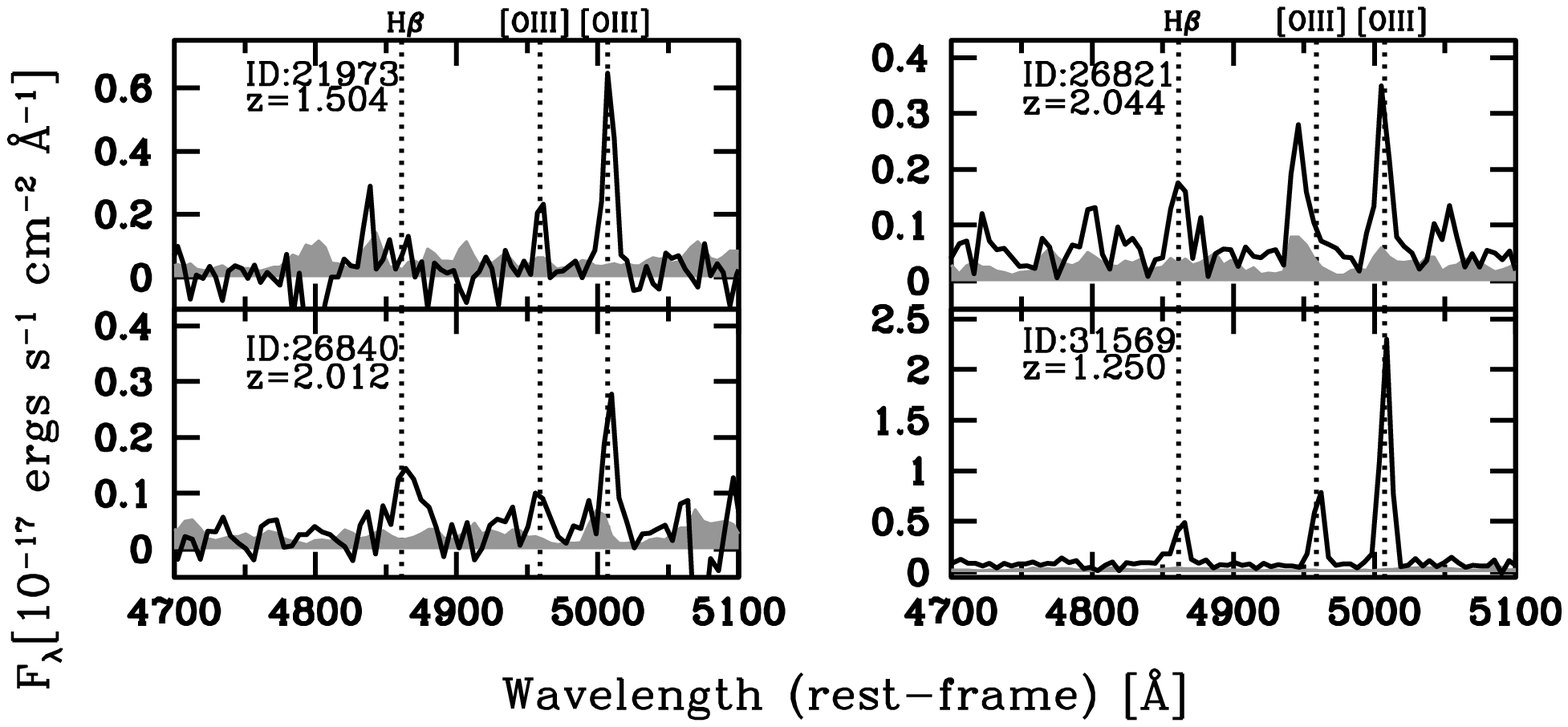}
  \caption{Same as Fig. \ref{fig;spectra-a}, but for $\Hb$ and [\ion{O}{3}].
           }
  \epsscale{1.0}
  \label{fig;spectra-b}
 \end{center}
 \begin{center}
  \epsscale{0.7}
  \plotone{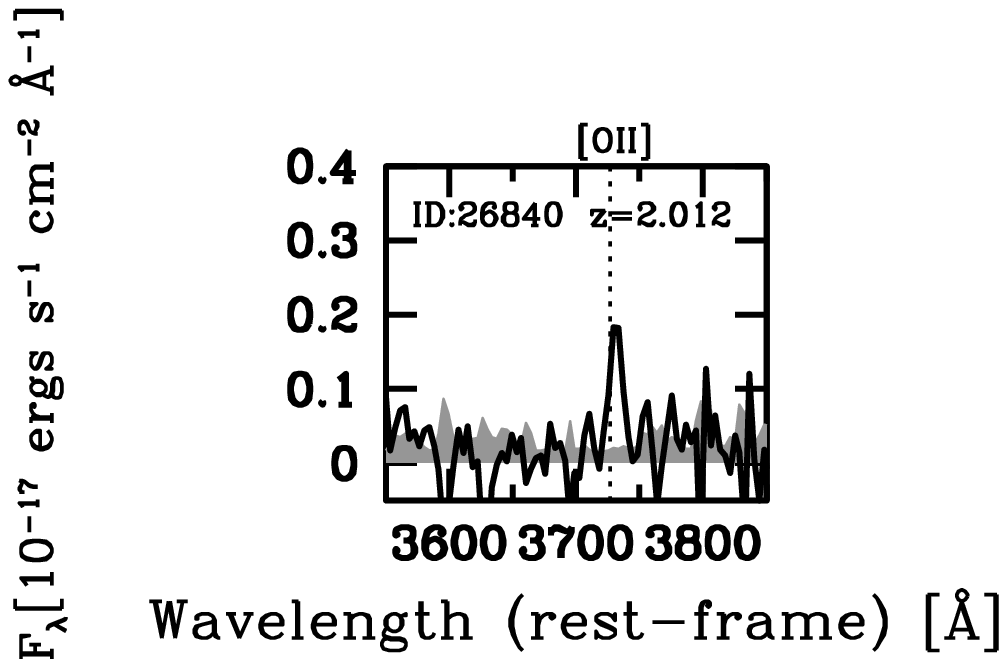}
  \caption{Same as Fig. \ref{fig;spectra-a}, but for [\ion{O}{2}].
           }
  \epsscale{1.0}
  \label{fig;spectra-c}
\end{center}
\end{figure}

We have detected $\Ha$ and some other emission lines in 15 of the 40
sBzKs, 9 of which have $K>22.0$.   
Since there are many strong OH airglow lines in the NIR region, we have
first confirmed all lines in the 2D spectra by visual inspection.
Among the confirmed lines, then, we regard lines with a peak flux larger
than the $3\sigma$ sky noise as a detection using 1D spectra with a
resolution of $R$=500, that is, unbinned spectra. 
For the 4 pBzKs, no spectra with enough S/N to detect any Balmer/4000\AA
\ break were obtained due to the insufficient exposure times. In what
follows, we examine and discuss spectroscopic properties of the 15 sBzKs 
with emission lines detected. Figures \ref{fig;spectra-a} --
\ref{fig;spectra-c} show the binned 1D spectra of the lines, and Tables
\ref{table;photo} and \ref{table;spec} show properties of the 15 sBzKs.  

To measure spectroscopic redshifts and fluxes of the emission lines,
Gaussian profiles with a constant continuum are fitted, 
where the free parameters are amplitude, line width, redshift, and
continuum level.    
The $1\sigma$ error of the spectrum is used as a sigma in the
calculation of $\chi^2$. 
Because $\Ha$ ($\lambda$6563) and [\ion{N}{2}]
($\lambda\lambda$6548,6584) are blended, they are simultaneously fitted,
assuming the same line width and redshift, and a 6584/6548 ratio of
3-to-1.  
The $1\sigma$ error of each parameter in the fitting is estimated as the
range where the increase in $\chi^2$ from the best-fit value is less
than the appropriate value for the number of degrees of freedom
(see \S14.5 of Numerical recipes by \citet{pre}).

\subsection{Spectroscopic redshifts}
\label{sec;spec-zflux}
Spectroscopic redshifts are derived from the $\Ha$ lines using the
central wavelength of the fitted Gaussian. For four sBzKs, which are
ID21360, ID21667, ID22602, and ID27677, only one emission line is
detected, but we regard it as $\Ha$ for the following reasons:  
(1) The redshift of ID21360 is also confirmed to be 1.174 by optical
spectroscopy with the Hectospec on the Multiple Mirror Telescope
\citep{ly08}.   
Therefore, the line is certainly $\Ha$.
(2) For the other three objects, it is possible that the detected lines, 
seen in the $HK$ grism spectra, are from transitions other than $\Ha$.
If the lines are $\Ha$, the redshifts should be 1.75, 1.87, and 1.71,
respectively. 
On the other hand, if the lines are [\ion{O}{3}]($\lambda$5007), those
would be 2.60, 2.76, and 2.55, respectively.
The transition of $\Hb$ would result in even higher redshifts.
Then, because the BzK technique isolates selecting star forming galaxies
at $1.4 \lesssim z \lesssim 2.5$ by detecting the Balmer break between
the $z$-band and $K$, $\Ha$ is the most likely identification of lines
in the $HK$ region. 
(3) In addition, we note that almost all the derived spectroscopic
redshifts are consistent with photometric redshifts.   
The photometric redshifts are obtained by fitting template SEDs to
$BVRi'z'K$ photometric data using the Hyper$z$ code (version 1.1)
\citep{bol00}. 
The used set of template SEDs is what is provided in the Hyper$z$
package, which consists of the observed spectra based on the CWW
templates \citep{cww80}. 
The standard deviation of $\sigma(z_{\rm spec} - z_{\rm phot})$ is 0.61 
for the 12 sBzKs with the secure spectroscopic redshfits. The
photometric redshifts for ID21667 and ID22602 are 1.47 and 1.46,
respectively, supporting our identification of their lines as $\Ha$. 
ID27677 would have a photometric redshift of 2.46, suggesting the
possibility that the line is $\Hb$ or [\ion{O}{3}]. 
Since the flux ratio of [\ion{O}{3}] to $\Hb$ can widely range from 0.1
to 10 \citep{vei87}, it is possible that only one of the two lines is
detected. 
Note that [\ion{O}{3}]($\lambda$4959) would not be detected due to the
inadequate S/N even if the identification of [\ion{O}{3}]($\lambda$5007) 
is correct. These considerations suggest that all the three of
$\Ha$, $\Hb$, and [\ion{O}{3}] are candidates for the single line of
ID27677, and that no secure conclusion is available for the line
identification.
Therefore, adopting the consideration (2), we assume that the line for
ID27677 is also most likely to be $\Ha$. Even if the line is not $\Ha$,
our conclusions discussed below would remain unchanged.

Since the spectral resolution is $R$=500, uncertainties in spectroscopic
redshifts are estimated to be
$\Delta z = \Delta\lambda_{\rm obs}(\Ha)/6563 \sim 0.002$. 
  
\begin{figure}[tb]
 \begin{center}
  \plotone{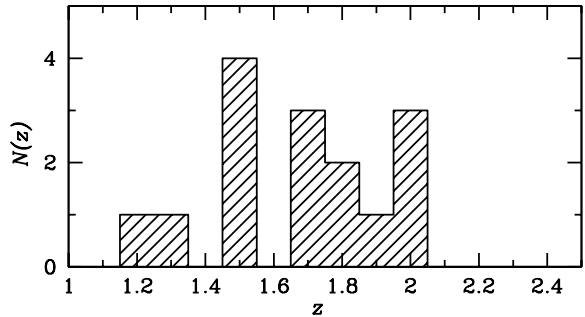}
  \caption{The redshift distribution of the sBzKs with spectroscopic
  redshift derived from the $\Ha$ lines. 
  Errors are $\Delta z \sim 0.002$.
           }
  \label{fig;Nz}
 \end{center}
\end{figure}

Figure \ref{fig;Nz} shows the redshift distribution of the sBzKs with
spectroscopic redshift, indicating a somewhat unexpected feature.
The distribution does not peak at $z \sim 2$, and no object lies at 
$z > 2$, contrary to the fact that $BzK$ color selection
is proposed to select galaxies at $1.4 < z < 2.5$. 
In fact, \citet{dad07} have reported that the redshift distributions of
sBzKs with spectroscopic and photometric redshifts both peak at 
$z \sim 2$, and that this selection technique works well to
select $z \sim 2$ galaxies, which is also supported by other results
with photometric redshifts \citep{gra07,bla08}.
The discrepancy of our result may be due to a flux-limited bias,
or the fact that the strongest emission line, H$\alpha$, is shifted
into a wavelength region of high sky background for $z > 2$, as
discussed in the next section.
The opposite problem also occurs, namely that some
galaxies at $z < 1.4$ are selected, which are also seen in Fig. 2 of
\citet{dad07}. These lower-$z$ objects are near the boundary of $BzK$
selection criterion, and lie 2.5 -- 2.7 sigma from the color selection
threshold. This result suggests that a fraction of star-forming galaxies 
at lower redshifts ($z < 1.4$) can meet the sBzK criterion.

\subsection{Objects without detected emission lines}
Among the sBzK targets, no emission line is detected for 25 sBzKs.
It is important to know the cause of the non-detections.
Judging from the distribution of these sBzKs without a line detection on
Figure \ref{fig;color_target}, it is possible that they have large dust
extinction.  Applying equation (4) of \citet{dad04} to the $B-z$ color,
we find that sBzKs with $E(B-V)$ larger than $\sim 0.5$--which
corresponds to Av=2.0 magnitude--have no lines detected. This may be one
of the reasons for the non-detection of lines. 
Since there are many strong OH airglow lines in the NIR region, it
is also possible that an emission line falls on a strong OH line. 
In addition, lines are redshifted into the $K$-band region in the case
of $z>2$ sBzKs. As described in section \ref{sec;spec-zflux}, the S/N in
the region may be too low for lines to be detected.    
In fact, there were no lines detected at a wavelength longer than
$2.0 \micron$. 
Finally, it is also possible that only sBzKs with higher SFRs have
strong enough lines to be detected in this $K$-limited sample.
The lowest $\Ha$ flux we detected is 
4.2 $\times 10^{-17}$ ergs s$^{-1}$ cm$^{-2}$.

\section{AGN contamination}
\label{sec;agn}
Before considering the spectroscopic properties of sBzKs, we should
account for possible AGN contamination in our sample. 
We use two methods to identify AGN candidates. 
One is the BPT emission-line diagnostic diagram \citep{bpt81}, and the
other is a power-law shaped SED from optical to MIR wavelengths. 

\begin{figure*}[tb]
 \begin{center}
  \plotone{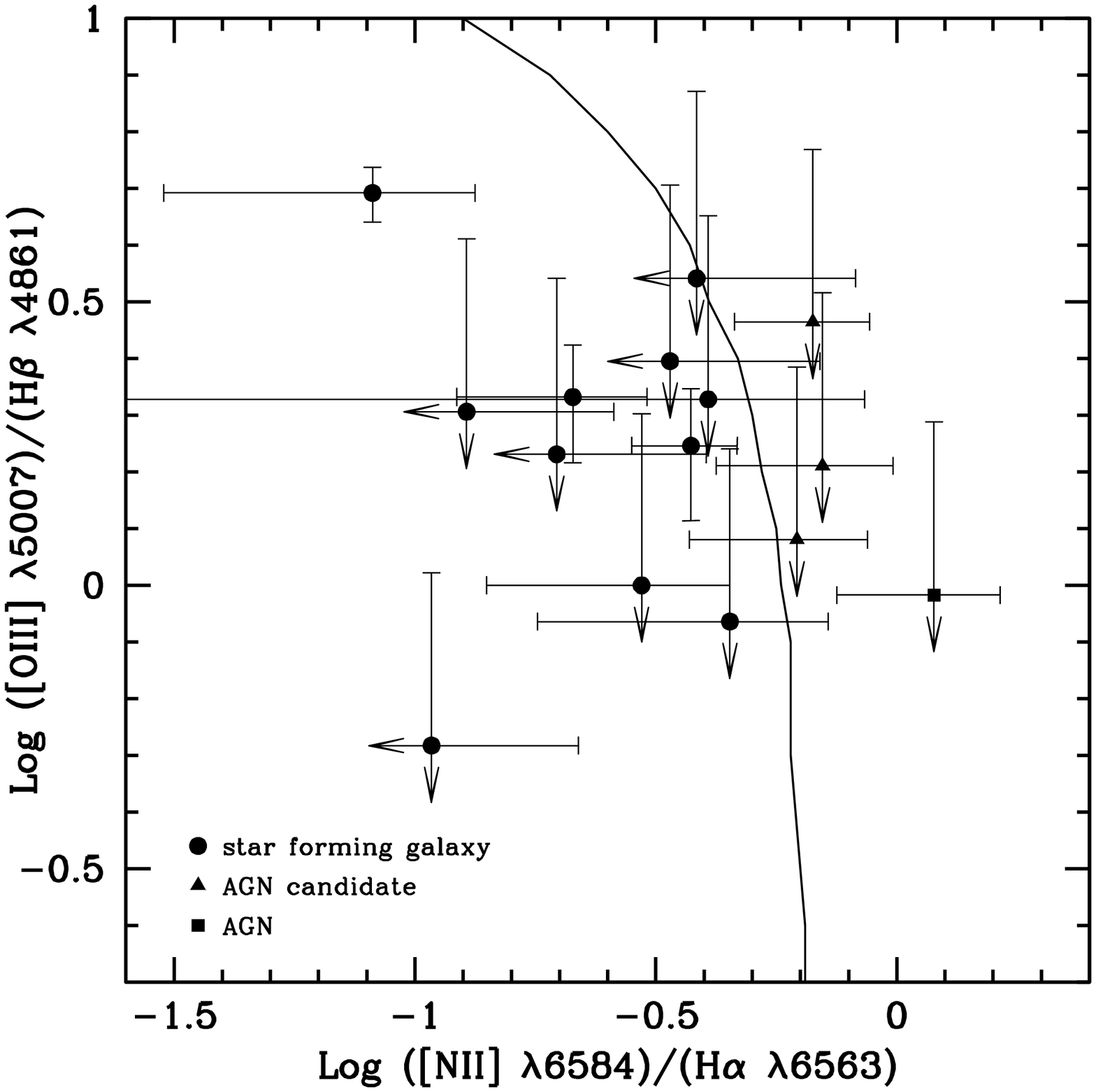}
  \caption{The emission-line diagnostic diagram showing the
  [\ion{O}{3}]($\lambda$5007)/$\Hb$
  vs. [\ion{N}{2}]($\lambda$6584)/$\Ha$ ratios. The solid line is the
  boundary separating star forming galaxies (left side) and AGNs (right
  side) \citep{kau03}. 
  Circles show objects identified as star forming galaxies, while a
  square and triangles as an AGN and possible AGN candidates,
  respectively.  
  The undetected fluxes of [\ion{O}{3}] and [\ion{N}{2}] are
  estimated from the $1\sigma$ upper limits, which are shown by
  arrows. The $\Hb$ fluxes are inferred from the $\Ha$ flux and $E(B-V)$
  on the assumption of an intrinsic Balmer decrement of
  $\Ha/\Hb=2.86$. The error bars are shown with $1\sigma$. 
  }
  \label{fig;BPT}
 \end{center}
\end{figure*}

Figure \ref{fig;BPT} is the emission-line diagnostic diagram showing the
[\ion{O}{3}]($\lambda$5007)/$\Hb$ vs. [\ion{N}{2}]($\lambda$6584)/$\Ha$
ratios. The only three sBzKs--ID26821, ID26840, and ID31569--have all
the required lines detected. Undetected line fluxes for the other
galaxies are estimated as follows. 
We derive $\Hb$ fluxes from the $\Ha$ fluxes, $E(B-V)$ described
in \S 5.1, and the assumed intrinsic Balmer decrement of $\Ha/\Hb=2.86$.   
The fluxes of [\ion{O}{3}] and [\ion{N}{2}] are the $1\sigma$ upper
limits derived from the sky noise at the wavelengths where each line
should be seen, assuming the same line width as $\Ha$.
As discussed in \citet{kew02} and \citet{erb06a}, the flux ratio of
[\ion{N}{2}]/$\Ha$ is enhanced by shock excitation (in LINERs), or by
the hard continuum which photo-ionizes the Narrow Line Region (NLR) of
all Seyfert nuclei. 
Figure 7 in \citet{kew02} shows that the ratio of [\ion{N}{2}]/$\Ha$ is
always less than 0.63 in purely star-forming galaxies.  
Among the 15 spectroscopically confirmed sBzKs in our sample, four--
ID21193, ID22424, ID25528, and ID31340--have [\ion{N}{2}]/$\Ha$ greater
than 0.63, which may imply a significant contribution from either a
LINER or Seyfert NLR component.
In particular, ID25528 shows the high [\ion{N}{2}]/$\Ha$ ratio of 1.2,
suggesting that it is an AGN. 
On the other hand, we cannot strongly assert that the other three
galaxies are LINERs or Seyferts due to the uncertainties of their line
ratios.
Also, as described below, the SEDs suggest that their continuum is
dominated by stellar emission, and that the AGN contribution is
small. If they are type II AGNs, the strongest observed contribution to
the spectrum from the active nucleus would be high-ionization emission
lines, not continuum.   
Similarly, one galaxy--ID 31569--shows a high enough
[\ion{O}{3}]($\lambda$5007)/$\Hb$ ratio of 4.9 to be a possible Seyfert  
2 galaxy. However, the weak [\ion{N}{2}] line suggests that this is
instead a metal-poor star-forming galaxy.

These ratios of forbidden to Balmer emission lines do not however
provide a definitive search for AGN. This is because a strong
contribution from the Broad Line Region (BLR) in Seyfert 1 nuclei
produces a low [\ion{N}{2}]/$\Ha$ ratio which would not be noticeably
different from those of \ion{H}{2} regions. 
Our spectral resolution of 600 km s$^{-1}$ should be sufficient
to detect most broad wing of $\Ha$, but no such wing is found.
Furthermore, only one of the $\Ha$ lines--ID 26821--satisfies the
criteria of \citet{hicks02} set for an Seyfert 1:
EW $>$ 100\AA \ and 
$\log L(\Ha,{\rm observed})[{\rm ergs\ s^{-1}}] \ge 42.5$ 
(Table \ref{table;spec}).  
This implies that there is a contribution from a Seyfert BLR component 
in ID26821 sBzK. However, as shown in \S \ref{sec;sfruv}, the SFR
derived from $\Ha$ luminosity is in agreement with that from UV
luminosity within a factor of 2, which may indicate that the
contribution from a BLR to the $\Ha$ line is not significant.

\begin{figure}[tb]
 \begin{center}
  \plotone{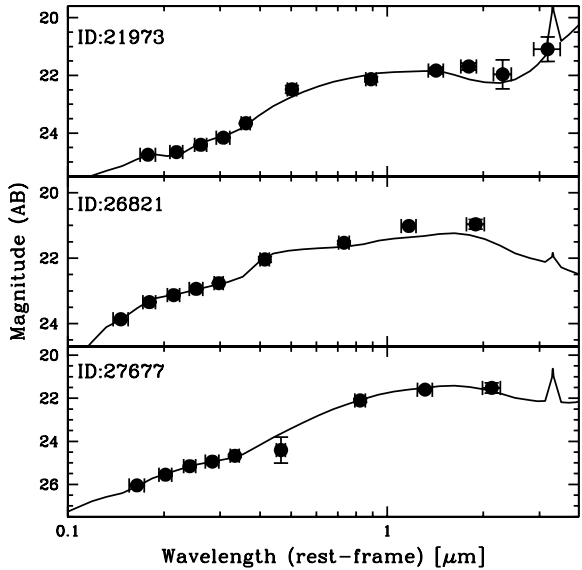}
  \caption{The long-wavelength power-law-like SEDs of ID21973, ID26821,
  and ID27677 sBzKs, where model SEDs of starbursts (without any AGN
  contribution) from \citet{takagi03} are overplotted.  
           }
  \label{fig;powerlaw}
 \end{center}
\end{figure}

We have also applied another test for Seyfert 1 nuclei in our sample. 
We checked the long-wavelength SEDs of the 15 sBzKs using the photometry
from $3.6 - 8.0 \mu m$ taken with IRAC on the Spitzer space telescope.  
Most of the sBzKs have a strong bump at a rest wavelength of 1.6 \micron
\ in their SEDs, indicating that their SEDs are dominated by stellar
emission \citep{saw02,alo07}. 
AGN-dominated galaxies, especially those with broad emission lines
(Seyfert 1's) have a power-law SED at rest wavelengths longer than
1 \micron \ \citep{mal83,spin95}.
In our sample, no such clear power-law SEDs are seen, but three
sBzKs have power-law-like SEDs: ID21973, ID26821, and ID27677 
(Figure \ref{fig;powerlaw}). 
These might perhaps contain "Seyfert 1.9" or "Narrow Line Seyfert 1"
nuclei. However, intense star formation also produces substantial
near-IR emission from heated dust, which could mimic such an SED. 
Figure \ref{fig;powerlaw} shows these three power-law-like SEDs with SED
models of starbursts overplotted \citep[SBURT:][]{takagi03}. 
These model SEDs have no contribution of AGN.

With all these considerations, we classify ID25528 and ID26821 as
AGN. Although the AGN contribution to the spectra, particularly the continuum,
may not be significant, the two galaxies are excluded from the
discussions on emission lines in the following section. 
We then classify the three sBzKs of ID21193, ID22424, and  ID31340 as AGN
candidates judging from the [\ion{N}{2}]/$\Ha$ ratios.
However, since it is possible that the three AGN candidates are star
forming galaxies, we consider them as well as the other ten sBzKs
classified as star forming galaxies in the following discussions.

\citet{kon06} have estimated that about 25\% of sBzKs are AGNs using
XMM-$Newton$ X-ray data, which would imply that 3 -- 4 sBzKs in our
sample could be AGN. However, the sample of \citet{kon06} has only
$K$-bright BzKs. According to \citet{alo07}, the characteristic stellar
mass of the host galaxy of AGN increases with redshift, and at 
$z \sim 2$ it is $\sim (1-3) \times 10^{11}\Msun$.   
In our sample, almost all galaxies are less than $10^{11}\Msun$, which
indicates that the fraction of AGN can be lower than 25\%.
The NICMOS imaging of \citet{colbert05} also indicates that fewer than a
quarter of these less massive galaxies should host an AGN detectable in
X-rays. 
The AGN fraction of 2/15 (13\%) in our sample is consistent
with the number of AGN expected from other studies.
If the three candidates are actually AGNs, the fraction increases to 33\%, 
which would still be consistent with other studies.

\section{Properties of star-forming BzK galaxies}
\label{sec;discussions}
Emission line information enables us to estimate SFR
and metallicity.  
In the local universe, more massive galaxies tend to have lower SFRs and
higher metallicities \citep[e.g.,][]{gal05}. 
It is important to examine whether a similar relation exists among these
properties for high redshift galaxies.

In this section, we first describe the properties obtained from
multiwavelength photometric data by SED fitting method. 
Then, spectroscopic properties are discussed, along with the
photometric properties.

\subsection{Properties from SED fitting}
\label{sec;sedfit}
Properties such as stellar mass and dust extinction, $E(B-V)$, are
obtained from fitting synthetic model spectra of continuous bursts
\citep{bru03} to the $BVRi'z'K$ photometry, assuming a standard \citet{sal55}
initial mass function (IMF), solar abundance, and the empirical dust
extinction of \citet{cal00}.   
These are shown in Table \ref{table;photo}.
Stellar mass is one of the most important parameters of galaxy properties
and the most robust one obtained from SED fitting.
Although sBzKs are well-fitted by the continuous burst model, fits using
the $\tau$ model, where the SFR declines exponentially with time, 
would make the stellar masses slightly smaller than those with the
continuous burst model.   
Note that SEDs of the simple stellar population (SSP) model are
irrelevant in this work, since sBzKs are still star-forming at 
$z \sim 2$.   
It is also noted that if a \citet{cha03} IMF is used, stellar masses
(and SFRs) are smaller by a factor of 1.8 than those with a Salpeter IMF. 
Thus, stellar masses are determined within a factor of $\sim 2.5$.
Fitted $E(B-V)$s are mainly used to calibrate the dust extinction of
emission line fluxes in section \ref{sec;sfr}. As discussed in section
\ref{sec;agn}, almost all SEDs of the sBzKs are dominated by stellar
emission with a clear 1.6$\mu m$ bump. Therefore, any AGN contribution
to the SED is small enough that parameters obtained from the SED fitting
reflect properties of the stellar populations.

\subsection{Star Formation Rates}
\label{sec;sfr}
\subsubsection{SFR derived from H$\alpha$ and UV continuum}
\label{sec;sfrha}
SFRs for sBzKs are derived from the luminosity of $\Ha$ using the equation
given in \citet{ken98};  
\begin{equation}
{\rm SFR}_{\Ha}[\Msun {\rm yr^{-1}}] = 7.9 \times 10^{-42} L_{\Ha}
[{\rm ergs\ s^{-1}}],
\end{equation}
assuming solar abundance, Salpeter IMF with mass limits of 0.1 and 100
$\Msun$, and a constant SFR.
$\Ha$ luminosities are corrected for the dust extinction by 
equation (3) in \citet{cal00}, 
$E(B-V) = 0.44 E_{\rm gas}(B-V)$.
The amount of nebular extinction is then estimated from 
$E_{\rm gas}(B-V)$ using the extinction law given in \citet{cal00}. 
The median amount of extinction of $\Ha$ is 2.83 magnitude. 

The amount of extinction at $\Ha$ can be directly estimated 
from the apparent $\Ha/\Hb$ ratio. However, we do not apply 
this method for our sample, because only three galaxies have 
detection of both lines and even for these three the amount 
of extinction is not strongly constrained due to a large 
uncertainty in the $\Hb$ flux.

The rest-UV luminosity is also used as a tracer of SFR. 
We derive SFR from the continuum at 1500\AA, $L_{1500}$, which
is predicted by the best-fit model to the $BVRi'z'K$ photometry.
$L_{1500}$ is corrected for the dust extinction with Calzetti extinction
law and converted into a SFR using the equation given in \citet{dad04};
\begin{equation}
{\rm SFR}_{\rm UV}[\Msun {\rm yr^{-1}}] = \frac{L_{1500}[{\rm ergs\
 s^{-1}\ Hz^{-1}}]}{8.85} \times 10^{-27}.  
\end{equation}

\subsubsection{Comparison between ${\rm SFR}_{\Ha}$ and ${\rm SFR}_{\rm UV}$}  
\label{sec;sfruv}
\begin{figure}[tb]
 \begin{center}
  \plotone{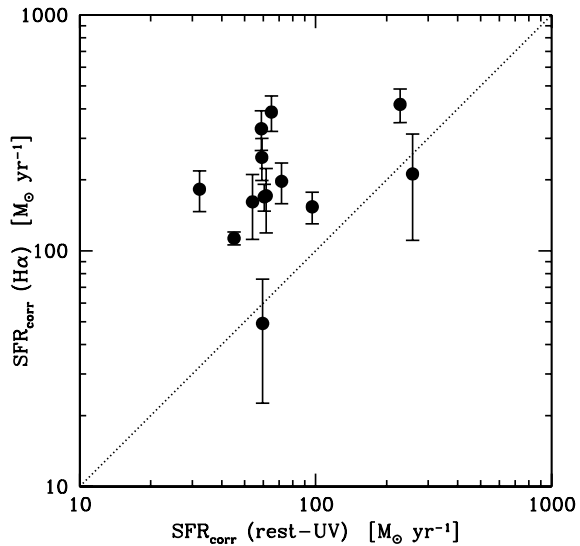}
  \caption{Comparison between ${\rm SFR}_{\Ha}$ and 
  ${\rm SFR}_{\rm UV}$. 
  ${\rm SFR}_{\Ha}$s are derived from $\Ha$ luminosities, and then 
  ${\rm SFR}_{\rm UV}$s are from the luminosities at 1500 \AA \  estimated
  from the best-fitted SEDs. 
  Both SFRs are corrected for dust extinction.
  The dotted line shows that ${\rm SFR}_{\rm UV}$ is equal to 
  ${\rm SFR}_{\Ha}$. 
           }
  \label{fig;sfr_ha_uv}
 \end{center}
\end{figure}

Figure \ref{fig;sfr_ha_uv} compares the SFR from the rest-UV
luminosity with that from $\Ha$. 
A correlation is seen between the two SFRs, but ${\rm SFR}_{\Ha}$ tends
to be larger than ${\rm SFR}_{\rm UV}$ by a factor of three. 
The median ${\rm SFR}_{\Ha}$ and ${\rm SFR}_{\rm UV}$ are 182 and
61 $\Msun {\rm yr^{-1}}$, respectively.  
Also, this discrepancy between SFR measured by $\Ha$ and from shorter
wavelength observations is quite similar to what \citet{hicks02} found
in their sample of IR-selected galaxies.

On the other hand, \citet{erb06b} report that ${\rm SFR}_{\Ha}$ for
UV-selected $z\sim 2$ galaxies is in good agreement with 
${\rm SFR}_{\rm UV}$, which is inconsistent with our result.
However, their observed $\Ha$ luminosities are similar to ours. That is,
the median $\Ha$ luminosities of our sample and that of \citet{erb06b}
are 2.0 and 2.2 times $10^{42}$ ergs s$^{-1}$, respectively. 
This disagreement on resulting SFRs with \citet{erb06b} may be due to 
different corrections for dust extinction at $\Ha$. They 
used the color excess for the stellar continuum to de-redden
$\Ha$, that is, $E_{\rm gas}(B-V)=E(B-V)$, which implies that
their amount of dust correction for $\Ha$ luminosity is systematically
lower than ours by 1.6 magnitude on average.  
This leads to a difference in derived SFRs of a factor of 4.4.
Using the same method as \citet{erb06b} for de-reddening $\Ha$,
our correlation between ${\rm SFR}_{\Ha}$ and ${\rm SFR}_{\rm UV}$
becomes consistent with that of \citet{erb06b}. 
In this case, the median ${\rm SFR}_{\Ha}$ is 50 $\Msun {\rm yr^{-1}}$.

\citet{dad04} report that most color excesses of 24 sBzKs in the
K20/GOODS area, which are derived from $U$ to $K$ SEDs, fall within a
range of $0.2 \lesssim E(B-V) \lesssim 0.5$, which is consistent with
our result (Table \ref{table;photo}). 
The median $E(B-V)$ in our sample is 0.38.
This implies that our estimation of $E(B-V)$ by SED fitting is reliable.

This difference between ${\rm SFR}_{\Ha}$ and ${\rm SFR}_{\rm UV}$
implies two possibilities. One is that a Salpeter IMF is unsuitable, if
the equation of \citet{cal00}, $E(B-V) = 0.44 E_{\rm gas}(B-V)$, 
is assumed to be valid for star-forming galaxies at $z \sim 2$. 
A more top-heavy IMF than Salpeter would account for the difference,
since nebular emission results from ionizing photons shortward of the
Lyman limit. 
The other is that the equation of \citet{cal00} cannot be applied to 
$z \sim 2$ galaxies. 
In any case, in what follows we use ${\rm SFR}_{\Ha}$ for sBzKs in
the comparison with their stellar masses.  
Even if ${\rm SFR}_{\rm UV}$ were used instead, the discussion in the next
section would remain unchanged.

\subsubsection{Relation to stellar mass}
\label{sec;RtoMstar}
\begin{figure}[tb]
 \begin{center}
  \plotone{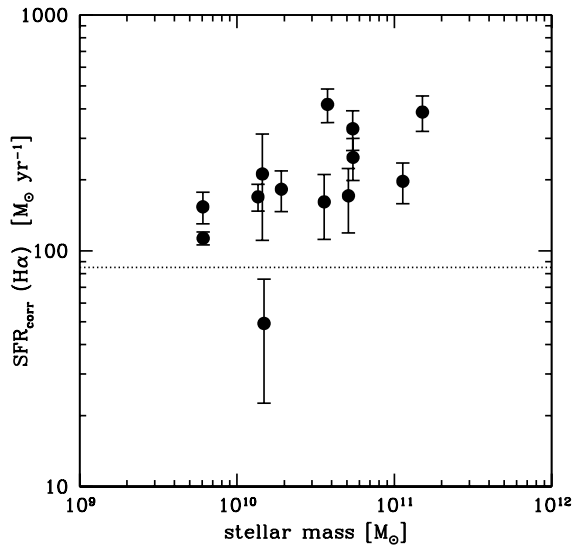}
  \caption{The relation between ${\rm SFR}_{\Ha}$ and stellar
  mass. Stellar masses are derived from SED fitting to $BVRi'z'K$
  photometry.    
  The dotted line shows the limiting SFR of 
  $87 \Msun {\rm yr^{-1}}$ estimated from the lowest detected flux on
  the assumption that the galaxy is at $<z>=1.7$ and suffer a typical
  color excess of 0.38. 
           }
  \label{fig;mstar_sfr}
 \end{center}
\end{figure}

We examine the relation between ${\rm SFR}_{\Ha}$ and stellar mass.
Figure \ref{fig;mstar_sfr} suggests that sBzKs with higher stellar
masses have larger SFRs, which is qualitatively consistent with
\citet{dad07}. 
However, the slope of our relation is shallower than theirs.
One cause for this disagreement in the slope may be a difference between
${\rm SFR}_{\Ha}$ and ${\rm SFR}_{\rm UV}$. 
Figure \ref{fig;sfr_ha_uv} suggests the difference between these two
estimates of SFR is larger at lower SFRs.
Since less massive sBzKs have lower SFRs, this difference results in
shallower slope of the correlation between stellar mass and 
${\rm SFR}_{\Ha}$. 
Other cause may be a difference in the method used to derive stellar
masses. \citet{dad07} have used the empirical relation between the
stellar mass and $K$ given in \citet{dad04}, which is calibrated based
on $z-K$ color. 
We compare the stellar masses derived from a SED-fitting with
those from the relation of \citet{dad04}, finding that the difference
between the two estimates of stellar mass is larger at smaller stellar
mass. The stellar masses from the relation are larger by a factor of 2
-- 3 in $\sim 10^{10} \Msun$. 
Because this relation is derived using bright $K<22$ BzKs,
the stellar masses of faint BzKs may be overestimated compared with
those from SED fitting, resulting in the shallower correlation
between stellar mass and ${\rm SFR}_{\Ha}$.

It is also possible that a flux-limited bias, discussed in \S\S 3.2 and
3.3, produces this disagreement.  
If galaxies with $\left<E(B-V)\right>=0.38$ and 
$F(\Ha)=4.2\times10^{-17} {\rm ergs\ s^{-1}\ cm^{-2}}$, which is the
lowest $\Ha$ flux we have detected, are at $\left<z\right>\sim1.7$, the
SFR would be $\sim87\Msun {\rm yr^{-1}}$. This suggests that it is
likely that we cannot detect any emission lines for galaxies with SFR
$\lesssim$ 87 $\Msun {\rm yr^{-1}}$. 
Moreover, it is found that ${\rm SFR}_{\Ha}$s is three times larger
than ${\rm SFR}_{\rm UV}$s on average. Combining this with the
relation between stellar mass and ${\rm SFR}_{\rm UV}$ given in Fig. 17
of \citet{dad07}, we find that the limiting SFR of 87 
$\Msun {\rm yr^{-1}}$ corresponds to a stellar mass of 
$\sim 1\times10^{10}\Msun$.  
These estimates indicate that it is possible that no galaxies with
$\lesssim 1\times10^{10}\Msun$ and $\lesssim 87 \Msun {\rm yr^{-1}}$ are
detected due to flux-limited bias. 
However, even if Figure \ref{fig;mstar_sfr} is limited to a region with
SFRs higher than the estimated detection limit, a mild correlation
between stellar mass and SFR can be seen.
If there are less massive galaxies with lower SFRs, as found in
\citet{dad07}, the correlation would be steeper.

\citet{erb06b} have also reported a similar correlation between SFR and
stellar mass for UV-selected $z \sim 2$ galaxies, but their correlation
is shifted to much lower SFR than ours. 
This is mainly due to differences in the assumed IMF as well as dust
correction, as discussed in section \ref{sec;sfruv}. 
In \citet{erb06b}, a Chabrier IMF is used to estimate stellar masses and
SFRs. SFRs derived with a Chabrier IMF are smaller than those with 
a Salpeter IMF by a factor of 1.8. 
These differences make our SFR 7.9 times larger than that of
\citet{erb06b}.
After correcting these systematic differences, the comparison of the
results between sBzKs and UV-selected galaxies indicates that the former  
tend to have slightly higher SFRs than UV-selected galaxies with the
similar stellar masses. This reflects the intrinsic difference
between the the dust amount in the two populations--only
sBzKs with higher SFRs may be easily detected, due to the larger dust
extinction than in UV-selected galaxies.

The positive correlation between SFR and stellar mass may be attributed
to a larger amount of gas in more massive galaxies, as suggested by
\citet{dad07}. 

\subsubsection{Specific SFR}
\label{sec;ssfr}
\begin{figure}[tb]
 \begin{center}
  \plotone{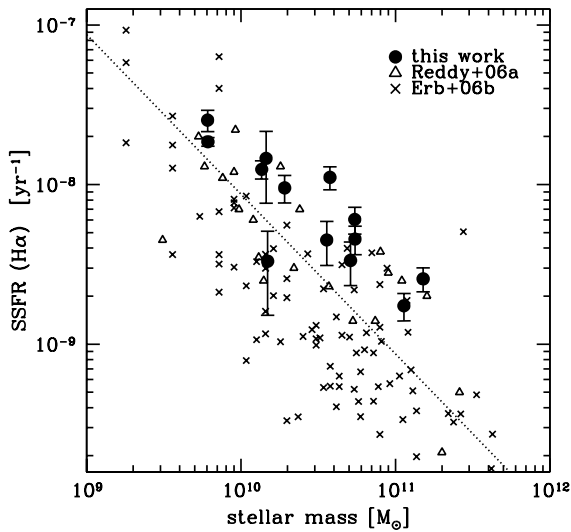}
  \caption{The relation between the specific ${\rm SFR}_{\Ha}$ and the
  stellar mass. 
  Filled circles are our results, open triangles show those of
  \citet{red06a} for sBzKs in their UV-selected sample, and crosses are
  those of \citet{erb06b} for UV-selected sample.
  All stellar masses and SSFRs are converted into those derived with
  Salpeter IMF and the same dust correction as we apply. 
  The dotted line shows the track of galaxies with a constant SFR
  of 87$\Msun {\rm yr^{-1}}$. 
           }
  \label{fig;mstar_ssfr}
 \end{center}
\end{figure}

We derived specific SFRs (SSFRs) from ${\rm SFR}_{\Ha}$s and stellar
Masses.  SSFR is an indicator of the time scale of star formation
activity. 
Figure \ref{fig;mstar_ssfr} shows SSFR as a function of stellar
mass. A negative relation is seen between stellar mass and SSFR, which
is consistent with previous results for $z \sim 2$ galaxies
\citep{red06a,erb06b}. 
This negative relation implies that a larger growth of stellar mass
occurs in less massive galaxies.  
In comparison with \citet{erb06b}, it should be noted that we apply
a different IMF and dust correction as described in the previous
section. In Figure \ref{fig;mstar_ssfr}, the results of \citet{erb06b}
are converted into those with the same manner as ours.

We must consider the effect of flux-limited bias on this correlation, as
discussed in the previous section.  
However, it seems that there is a genuine negative correlation even if
the range of validity is limited to stellar masses larger than
$1\times10^{10}\Msun$.  
The steep relation between stellar mass and SSFR results from the
shallow correlation between stellar mass and SFR, which shows that the
increase in SFR is not as rapid as that in stellar mass. 
If we miss galaxies with low SFRs of $\lesssim87\Msun {\rm yr^{-1}}$ as
discussed in section \ref{sec;RtoMstar}, that is, the correlation
between stellar mass and SFR is steeper, the relation with SSFR would be
shallower than that in Figure \ref{fig;mstar_ssfr}.   

\subsection{Equivalent Width}
\label{sec;ew}
\begin{figure}[tb]
 \begin{center}
  \plotone{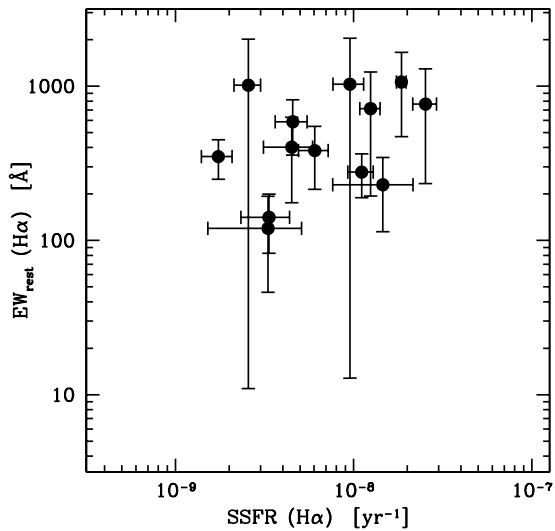}
  \caption{The relation between specific star formation rate and
  $\Ha$ equivalent width in rest frame. Rest-frame equivalent widths of
  $\Ha$ are derived from line and continuum fluxes after correction for
  dust extinction. 
           }
  \label{fig;ew_sfr}
 \end{center}
\end{figure}

The rest-frame equivalent widths of $\Ha$ are derived from 
Gaussian fitting. 
Both $\Ha$ and continuum fluxes have been corrected for dust
extinction.
As discussed in \citet{erb06b}, one can infer the star formation history
of galaxies from the equivalent width. 
While the flux of $\Ha$ reflects the current star forming activity,
the continuum flux emitted from stars reflects the past star
formation. That is to say, the equivalent width represents how 
actively a galaxy is forming stars compared with the past.
Thus, the equivalent width is a similar indicator to SSFR.
Figure \ref{fig;ew_sfr} shows the equivalent width as a function of the
SSFR, indicating a mild correlation. 
This supports the result of the previous section. 
Less massive galaxies are more actively forming stars, which results in
a larger growth of stellar mass.

\subsection{Metallicity}
\label{sec;metallicity}
Average metallicity is one of the most important properties of a galaxy. 
Since gas and stellar metallicities are related to the past star
formation, they are crucial for understanding the star formation history
and galaxy evolution. 
It is well-known that metallicity is correlated with stellar mass in the
local universe \citep[e.g.,][]{tre04}. 
\citet{erb06a} have reported a similar correlation to the local one
for UV-selected galaxies at $z \sim 2$.  However, their mass-metallicity
($M$-$Z$) relation at $z \sim 2$ is shifted to lower metallicity by 
$\sim 0.3$ dex. 
This $M$-$Z$ relation at $z \sim 2$ obtained by \citet{erb06a} is widely
used as representative of $z \sim 2$ star forming galaxies
in comparison with both observational and theoretical studies
\citep{liu08,mai08,bro07,der07,fin08}. However, because the UV selection
misses star forming galaxies with a large dust extinction, we cannot
be confident that the $M$-$Z$ relation of \citet{erb06a} 
applies to all star-forming galaxies at $z \sim 2$. 
Here, we determine a $M$-$Z$ relation for sBzKs.
Our sBzK sample, which approximately corresponds to a stellar 
mass-limited sample, should be more suitable for deriving the $M$-$Z$
relation at $z \sim 2$. 

It is well known that electron temperature reflects
gas metallicity \citep{kew02,kob04,erb06a}. 
Because auroral lines, which are used to derive electron temperature,
are weak, and observed in only galaxies with low metallicities, 
it is very difficult even in local galaxies to estimate the metallicity
of galaxies from them.
We can only use the ratio of strong emission lines which are emitted
from different ionization levels to derive metallicity for high-$z$
galaxies. At present, there are various metallicity diagnostics, such as
the $R_{23}$ and N2 methods, to estimate gas-phase oxygen abundance,
12+log(O/H). One problem with these diagnostics is a disagreement
between the resulting oxygen abundances inferred from the different
diagnostics. Even if the same diagnostic is used, some different
calibrations give a large difference in the resulting metallicities
\citep{kew02,kob04}. The same diagnostics and calibration should be used
to properly compare with other results. 

We derive 12+log(O/H) using the N2 index method proposed by \citet{sto94},
where N2 is defined as the flux ratio of [\ion{N}{2}]($\lambda6584$)/$\Ha$;
\begin{equation}
{\rm N2} \equiv \log\frac{[{\rm NII}](\lambda6584)}{\Ha}.
\end{equation}
The gas-phase oxygen abundances are derived with the following relation
given in \citet{pet04};
\begin{equation}
\begin{array}{l}
12+\log({\rm O/H}) = 8.90 + 0.57\times{\rm N2}.
\end{array}
\end{equation}
The reasons we use this relation are that it requires only two
lines of $\Ha$ and [\ion{N}{2}], and the result can be compared
directly with that of \citet{erb06a}, 
where the same relation was used to derive the gas abundance.    
Also, it should be noted that the lines of $\Ha$ and [\ion{N}{2}] are close
enough that the difference in dust extinction is negligible. 
The $1\sigma$ dispersion in the abundance derived from the relation is
0.18 \citep{pet04}.

\begin{figure*}[htb]
 \begin{center}
  \plotone{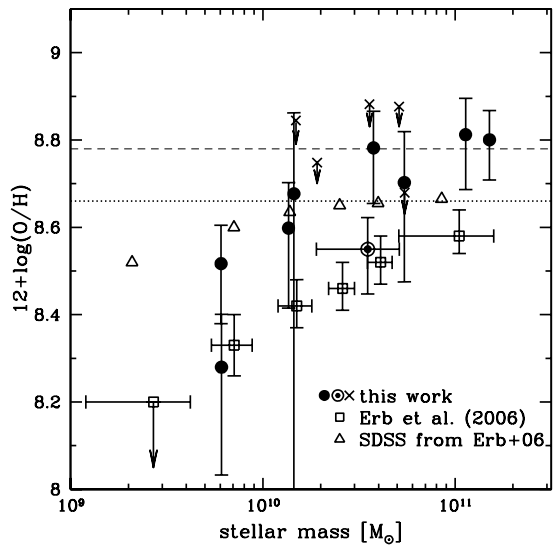}
  \caption{Stellar mass vs. metallicity relation. Circles and crosses
  show our results. Filled circles show those of sBzKs with the detection
  of both $\Ha$ and [\ion{N}{2}]. 
  Crosses show upper limits of sBzKs without the detection of [\ion{N}{2}],
  which are derived using the $3\sigma$ upper limit on [\ion{N}{2}] flux. 
  The double circle shows the mean abundance of the five sBzKs without a
  detection of [\ion{N}{2}] against the mean stellar mass.
  Open squares show UV-selected $z \sim 2$ galaxies, and
  open triangles are those of the local $\sim 53,000$ SDSS galaxies 
  \citep{erb06a}. All abundances are estimated using the same N2 index
  method. The dotted line is the solar oxygen abundance of 
  12+log(O/H)=8.66 \citep{asp04}.
  The broken line shows the abundance for galaxies with the flux ratio
  of [\ion{N}{2}]/$\Ha$=0.63.  
  Stellar masses of the \citet{erb06a} are converted into those derived 
  with Salpeter IMF. 
  } 
  \label{fig;mstar_metal}
 \end{center}
\end{figure*}

Figure \ref{fig;mstar_metal} shows the resulting relation between
stellar mass and metallicity. 
The broken line shows the abundance for galaxies with the flux ratio of
[\ion{N}{2}]/$\Ha$=0.63, where N2 saturates \citep{kew02}, suggesting
that the linear relation between the oxygen abundance and N2 index is
acceptable up to the abundance of $\sim8.79$.  
As discussed in section \ref{sec;agn}, the three sBzKs with a large
abundance, ID21193, ID22424, and ID31340, may harbor AGNs, because they
have flux ratios of [\ion{N}{2}]/$\Ha$ larger than 0.63. However, their SEDs
suggest that their continuum is dominated by stellar emission, and that
the AGN contribution is small. 
Even if the three AGN candidates are excluded from our sample, our 
conclusions in the following discussion are unchanged.    

We derive upper limits of metallicity for sBzKs without a secure
detection of [\ion{N}{2}], assuming that the peak flux is 3 times sky noise,
and that the line width is the same as that of $\Ha$. 
These galaxies are shown by crosses in the figure.
Also, we estimate mean metallicity of the five sBzKs without
[\ion{N}{2}] detection. We derive it from the stacked spectrum using the
equations (3) and (4), finding that the mean abundance is 8.55.
This is shown by the double circle in the figure, whose stellar mass is the
average of those of the five sBzKs.

Figure \ref{fig;mstar_metal} shows that more massive sBzKs tend to be
more metal-rich, a trend qualitatively consistent with that of
\citet{erb06a}. 
However, the metallicities of sBzKs with [\ion{N}{2}] detection are
higher by $\sim 0.2$ dex than those of UV-selected galaxies with the
same stellar mass at the similar redshift. Considering the
uncertainties, a significant shift exists at the 2.7$\sigma$ level
between the two $M$-$Z$ relations.  
If three AGN candidates are excluded from the sample, the confidence
level in the shift slightly decreases, to the 1.8$\sigma$ level.

We again emphasize that the N2 flux ratios are converted into the oxygen
abundances using the same relation as \citet{erb06a}.
We make sure that the N2 indices are larger than those of
\citet{erb06a} for galaxies with the given stellar masses, suggesting
that the cause of the difference in $M$-$Z$ relation is not a systematic
difference in our abundance estimate methodology. 

\citet{pet04} have found a relation between gas oxygen abundance and
O3N2 index, which is defined as 
${\rm O3N2}$ $\equiv$
$\log$[([\ion{O}{3}]($\lambda5007$)/$\Hb$) /
([\ion{N}{2}]($\lambda6584$)/$\Ha$)].  
We check the abundance with the O3N2 diagnostics for three sBzKs with
the necessary lines observed. The abundances from the O3N2 ratios are
lower than those from the N2 ratios by $\sim 0.13$ dex. This is
consistent with \citet{erb06a}, who found that the abundances from O3N2
are 0.17 dex lower on average than those from N2.  
In addition, it is not likely that the systematic difference in
estimates of stellar masses accounts for the difference in the $M$-$Z$
relation.   
The stellar masses of \citet{erb06a} are converted into those derived
with Salpeter IMF. 
As discussed in section \ref{sec;sedfit}, the systematic difference of
stellar masses should be less than a factor of 2.
This indicates that there is no way that it is the cause of the
difference in $M$-$Z$ relation. 

We consider the possibility that our $M$-$Z$ relation
suffers from a selection bias that only sBzKs with high metallicities
are plotted, since metallicities are properly measured only for sBzKs
with both $\Ha$ and [\ion{N}{2}] lines detected.   
The stacking analysis for sBzKs without [\ion{N}{2}] detection
suggests that there are also sBzKs with lower metallicities
(Figure \ref{fig;mstar_metal}), supporting the idea that
selection bias cannot be completely ruled out.
This may imply that the mean metallicities of sBzKs with given
stellar masses are slightly lower than those plotted in Figure 
\ref{fig;mstar_metal}. 
However, the mean metallicity of only sBzKs without [\ion{N}{2}]
detection is similar to that of UV-selected galaxies, implying that
sBzKs would still be more metal rich on average than UV-selected
galaxies. This fact suggests that it is likely that the UV selection
misses star forming galaxies at $z\sim2$ with comparatively higher
metallicities, and that metallicities of star forming galaxies at 
$z \sim 2$ have a large dispersion.

Therefore, we conclude that the oxygen abundance of sBzKs is greater
than that of UV-selected $z \sim 2$ galaxies. 
As described in section \ref{sec;sfr}, $K$-selected star forming
galaxies at $z\sim2$ have slightly higher SFRs but similar to
UV-selected galaxies, if the same correction for dust extinction is
applied, while their metallicities are significantly different.   
This fact may imply that the two star forming galaxy populations at 
$z \sim 2$ pass through different star forming histories before this
epoch.

\begin{figure}[tb]
 \begin{center}
  \plotone{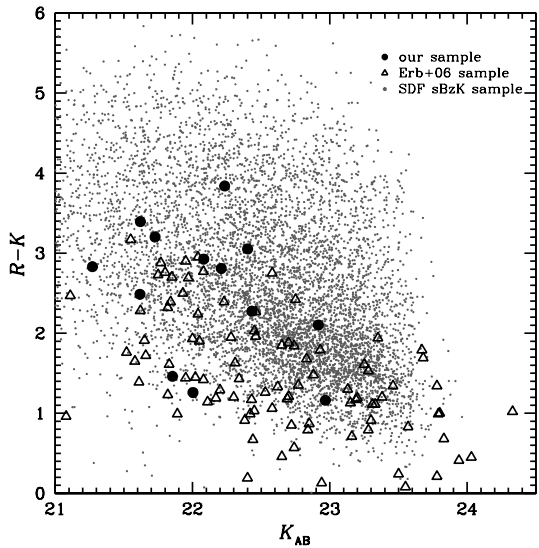}
  \caption{The color magnitude diagram of $R-K$ vs. $K$ for our sBzK
  spectroscopic sample (filled circles), the whole sBzK sample in the SDF
  (dots) and the NIRSPEC sample of \citet{erb06c} (open triangles).  
           }
  \label{fig;k_rk}
 \end{center}
\end{figure}

This difference in the $M$-$Z$ relation between sBzKs and UV-selected
galaxies may be related to the difference in color of the two
populations. 
Figure \ref{fig;k_rk} shows $R-K$ color as a function of $K$ for our
sBzK sample and the UV-selected galaxy sample of \citet{erb06a}.
This figure shows that the sBzK spectroscopic sample galaxies are
redder than the UV-selected galaxies.
Indeed, the sBzK ID31569, which is as blue as UV-selected galaxies, is
located on the $M$-$Z$ relation of \citet{erb06a}. The difference in
$R-K$ color may imply that galaxies with different colors have different
ages and/or dust contents, since the 4000\AA \ break falls
between the $R$ and $K$-bands. This fact indicates that star forming
galaxies at $z \sim 2$ with redder colors have larger metal abundances,
and that the $M$-$Z$ relation of \citet{erb06a} may not represent the
entire population of $z \sim 2$ galaxies.

Another possibility is that the difference in redshifts of galaxies
between the two samples accounts for the difference in the $M$-$Z$
relation.  
While the median redshift of galaxies in our sample is 1.7, the mean
redshift is $\left<z\right>=2.2$ in the \citet{erb06a} sample. Because the
galaxies in our sample are on average at a lower redshift, the fact that 
our sBzKs have higher metallicities than UV-selected galaxies of
\citet{erb06a} may plausibly be attributed to cosmic evolution.  
However, this interpretation would require a surprisingly rapid
evolution of the $M$-$Z$ relation over 800 Myr.

We plot results of star forming galaxies at $z \sim 0$ from
SDSS data in Figure \ref{fig;mstar_metal} \citep{tre04,erb06a}. It
should be noted that we cannot directly discuss redshift evolution of
the $M$-$Z$ relation, since these SDSS galaxies are not necessarily
descendants of sBzKs.    
Our $M$-$Z$ relation seems to be similar to that of SDSS galaxies
\citep{tre04}, but we may not necessarily claim that the metallicities
of sBzKs are the same as those of the local SDSS galaxies.  
There is the possibility that the true metallicities of the SDSS
galaxies are higher than those derived by the N2 method, if the
saturation of the N2 index at high metallicity is taken into 
account. Thus, the oxygen abundances of sBzKs can be as high as those of
the SDSS galaxies.  
Then, our result that the metallicities of sBzKs are close to solar
abundance is consistent with the assumption of the SEDs with solar
abundance in SED fitting.

It is also found that there are $M$-$Z$ correlations at other
redshifts. 
\citet{sav05} have reported the relation for NIR-selected
galaxies at $z\sim0.7$, finding that the metallicities of galaxies with
a given stellar mass have a dispersion of $\sim0.2$ dex around the mean 
abundances. The correlation with large scatter is similar to what we
found for galaxies at $z\sim2$.
\citet{mai08} have found the relation shifted to much lower metallicity
for 9 Lyman Break Galaxies at $z\sim3.5$, suggesting that the
metallicity of galaxies decreases with increasing redshift. 
However, since metallicity measurements for high-$z$ galaxies are very
limited, especially at $z>2$, the $M$-$Z$ relations of high-$z$ galaxies
obtained so far may have significant statistical and systematic
uncertainties. Therefore, in future, it is essential to derive more
reliable $M$-$Z$ relations at high redshifts from a large, unbiased
sample. 

\section{Summary}
\label{sec;conclusion}
We have carried out NIR spectroscopy of 44 BzK galaxies with 
$\KAB \lesssim 23$ in the Subaru Deep Field with MOIRCS on the Subaru
Telescope.  The $zJ$ and
$HK$ grism spectra cover wavelengths from 0.9 to 2.3 $\micron$ at a
resolution of $R$=500. 
The 44 spectroscopic targets are selected from a large sample of BzKs in
180 arcmin$^2$ of the SDF. 
Among the 44 targets, 40 are classified as sBzKs and 4 as pBzKs. Half
of the 44 targets are fainter than $\KAB=22$. We detect $\Ha$ and some other
emission lines in 15 sBzKs, nine of which have $\KAB > 22$.  

Spectroscopic redshifts are derived from $\Ha$, and fluxes
of emission lines are measured.
Stellar mass and dust extinction are also obtained from SED-fitting to
multiwavelength imaging data of $BVRi'z'K$.    
We have then obtained the following results.

The spectroscopic redshift distribution of the sBzKs does
not peak at $z \sim 2$, and objects at $z < 1.4$ are also
selected. These lower-$z$ objects are near the selection boundary 
in the $BzK$ diagram. On the other hand, no $z > 2$ objects are
detected. This result suggests that a fraction of star forming galaxies
at low redshifts can meet the sBzK criterion. However, the lack of
objects at $z > 2$ may be due to a flux-limited bias. 

The $\Ha$ luminosities are converted into SFRs. 
It is found that sBzKs with higher stellar masses have larger SFRs.
The correlation between SFR and stellar mass may be attributed to a
larger amount of gas for more massive galaxies.
Dividing SFR by stellar mass, specific SFRs are also derived.
A negative relation is seen between stellar mass and SSFR, which is
consistent with previous results for $z \sim 2$ galaxies. 
This negative relation implies that a proportionately larger growth of stellar
mass occurs in less massive galaxies. 

Gas-phase oxygen abundances, 12+log(O/H), are derived from the ratio of
[\ion{N}{2}]($\lambda6584$) to $\Ha$ using the N2 index method. 
We have found a correlation between stellar mass and oxygen abundances,
in the sense that more massive sBzKs tend to be more metal-rich, which
is qualitatively consistent with the relation for UV-selected $z \sim 2$
galaxies. However, the metallicities of sBzKs are on average higher by
$\sim 0.2$ dex than that of UV-selected galaxies at similar redshifts,
and are close to those of local galaxies. 
Compared with the uncertainties, this difference in metallicity between
sBzKs and UV-selected galaxies is significant, and may be due to the
color difference between the two populations. 
Our sBzKs have redder $R-K$ colors than the UV-selected
galaxies. Further, an sBzK with a blue color similar to the UV-selected
galaxies has a lower abundance, comparable to those of the UV-selected
galaxies. These facts suggest that there is a color-dependence in the
oxygen abundance in galaxies at $z \sim 2$, which may imply that
galaxies with different ages and/or dust amounts have different
metallicities.

\acknowledgments
We would like to thank the Subaru Telescope staff for their invaluable
help in our observation with Subaru/MOIRCS. 
We also thank an anonymous referee for useful comments, which have
greatly improved the paper.
M. Hayashi acknowledges support from the Japan Society for the Promotion
of Science (JSPS) through JSPS Research Fellowship for Young Scientists.

\clearpage
\begin{deluxetable}{cccccc}
\tabletypesize{\small}
\tablewidth{0pt}
\tablecolumns{6}
\tablecaption{Photometric properties of 15 sBzKs with emission lines detected. \label{table;photo}}
\tablehead{
 \colhead{} &  Magnitude &
 \multicolumn{2}{c}{Color} & \multicolumn{2}{c}{SED Fitting}            \\  
 \colhead{ID}& \colhead{$K$} & \colhead{$B-z$} & \colhead{$z-K$} &
 \colhead{stellar mass\tablenotemark{a}}& \colhead{$E(B-V)$}\\}

\startdata
21193&  21.3& 1.69& 1.89& 3.8 & 0.51\\ 
21360&  21.6& 1.61& 1.53& 1.5 & 0.43\\
21667&  22.1& 1.24& 2.15& 5.5 & 0.38\\
21973&  22.4& 1.09& 1.53& 1.9 & 0.31\\
22424&  22.2& 1.72& 2.65& 15  & 0.45\\
22602&  22.2& 1.23& 1.98& 5.1 & 0.35\\
25528&  22.3& 1.21& 1.49& 1.2 & 0.41\\
26411&  21.7& 1.47& 2.38& 5.4 & 0.44\\
26821&  21.8& 1.10& 1.24& 2.1 & 0.38\\
26840&  23.0& 0.73& 0.89& 0.61& 0.29\\
27677&  22.4& 1.37& 2.57& 3.6 & 0.45\\
27815&  22.9& 1.06& 1.45& 1.4 & 0.36\\
28761&  21.9& 1.04& 1.02& 1.5 & 0.35\\
31340&  21.6& 1.46& 2.53& 11  & 0.41\\
31569&  22.0& 0.78& 0.74& 0.61& 0.23\\
\enddata

\tablenotetext{a}{The unit is $10^{10} \Msun$.}

\end{deluxetable}

\clearpage
\begin{deluxetable}{cccccccc}
\tabletypesize{\small}
\tablewidth{0pt}
\tablecolumns{8}
\tablecaption{Spectroscopic properties of 15 sBzKs with emission lines detected.\tablenotemark{a} \label{table;spec}}
\tablehead{
 \colhead{ID} & \colhead{Detected Lines\tablenotemark{b}}&
 \colhead{redshift}& \colhead{$F_{\Ha}$\tablenotemark{c}} &
 \colhead{$L_{\Ha}$\tablenotemark{d}} &
 \colhead{SFR$_{\Ha}$\tablenotemark{e}} &
 \colhead{EW$_{\Ha}$\tablenotemark{f}} &
 \colhead{12+log(O/H)\tablenotemark{g}}\\}

\startdata
21193& $\Ha$,[\ion{N}{2}]& 1.479& 
     10.9$\pm$1.77& 1.5$\pm$0.24& 4.2$\pm$0.68& 1.8$\pm$0.56&
8.78$^{+0.08}_{-0.13}$\\ 

21360& $\Ha$      & 1.174& 
     4.16$\pm$2.25& 0.32$\pm$0.18& 0.49$\pm$0.27& 0.86$\pm$0.53&
$<$8.84\\

21667& $\Ha$      & 1.750& 
     11.1$\pm$2.26& 2.3$\pm$0.47& 2.5$\pm$0.51& 4.5$\pm$1.8&
$<$8.68\\

21973& $\Ha$,[\ion{O}{3}]& 1.504& 
     18.3$\pm$3.58& 2.6$\pm$0.51& 1.8$\pm$0.36& 8.5$\pm$8.4&
$<$8.75\\

22424& $\Ha$,[\ion{N}{2}] & 2.018& 
     7.25$\pm$1.24& 2.1$\pm$0.37& 3.9$\pm$0.66& 7.0$\pm$6.9&
8.80$^{+0.07}_{-0.09}$\\

22602& $\Ha$       & 1.872& 
     7.72$\pm$2.35& 1.9$\pm$0.58& 1.7$\pm$0.52& 1.1$\pm$0.46&
$<$8.88\\

25528& $\Ha$,[\ion{N}{2}] & 1.506& 
     9.99$\pm$2.00& 1.4$\pm$0.29& 2.0$\pm$0.40& 3.3$\pm$1.4&
8.94$^{+0.08}_{-0.12}$\\

26411& $\Ha$,[\ion{N}{2}] & 1.498& 
     14.0$\pm$2.67& 2.0$\pm$0.38& 3.3$\pm$0.63& 2.7$\pm$1.2&
8.70$^{+0.12}_{-0.23}$\\

26821& $\Ha$,$\Hb$,[\ion{O}{3}],[\ion{N}{2}]& 2.044& 
     22.2$\pm$1.43& 6.8$\pm$0.44& 7.3$\pm$0.47& 5.9$\pm$1.2&
8.66$^{+0.06}_{-0.07}$\\

26840& $\Ha$,$\Hb$,[\ion{O}{2}],[\ion{O}{3}],[\ion{N}{2}]& 2.012& 
     8.98$\pm$1.37& 2.6$\pm$0.40& 1.5$\pm$0.23& 6.6$\pm$4.6&
8.52$^{+0.09}_{-0.14}$\\

27677& $\Ha$       & 1.710& 
     4.53$\pm$1.39& 0.89$\pm$0.27& 1.6$\pm$0.50& 2.8$\pm$1.6&
$<$8.88\\

27815& $\Ha$,[\ion{N}{2}] & 1.796& 
     7.75$\pm$1.01& 1.7$\pm$0.22& 1.7$\pm$0.22& 5.5$\pm$4.0&
8.60$^{+0.10}_{-0.18}$\\

28761& $\Ha$,[\ion{N}{2}] & 1.818& 
     10.3$\pm$4.90& 2.3$\pm$1.1& 2.1$\pm$1.0& 1.8$\pm$0.91&
8.68$^{+0.19}_{-8.68}$\\

31340& $\Ha$,[\ion{N}{2}] & 1.719& 
     7.10$\pm$1.39& 1.4$\pm$0.28& 2.0$\pm$0.39& 2.5$\pm$0.73&
8.81$^{+0.08}_{-0.13}$\\

31569& $\Ha$,$\Hb$,[\ion{O}{3}],[\ion{N}{2}]& 1.250& 
     33.0$\pm$2.05& 3.0$\pm$0.19& 1.1$\pm$0.07& 9.9$\pm$5.5&
8.28$^{+0.12}_{-0.25}$\\
\enddata

\tablenotetext{a}{
 For the two sBzKs classified as an AGN and excluded
 from the discussions in \S 5, the SFR, equivalent width, and oxygen
 abundance derived by the same manners as the other sBzKs are also
 listed.}   
\tablenotetext{b}{[\ion{N}{2}] means [\ion{N}{2}]($\lambda6584$).}
\tablenotetext{c}{Observed flux. The unit is $10^{-17}$ ergs s$^{-1}$ cm$^{-2}$.}
\tablenotetext{d}{Observed luminosity. The unit is $10^{42}$ ergs s$^{-1}$.}
\tablenotetext{e}{The unit is $10^2 \Msun {\rm yr^{-1}}$. Dust-corrected.}
\tablenotetext{f}{The unit is $10^2$ \AA. Rest-frame and dust-corrected.}
\tablenotetext{g}{12+log(O/H) is derived from N2 index method.}
\end{deluxetable}

\end{document}